\documentclass[a4paper,twoside,12pt]{article}

\usepackage{amsmath}
\usepackage{amssymb}
\usepackage{latexsym}
\usepackage{cite}
\usepackage{graphicx}
\newcommand{\gthree}{{\gamma_\ast}}

\newcommand{\medsp}{\\[0.7ex]}

\newcommand{\tve}{\underline{\varepsilon}}

\newcommand{\ve}{\varepsilon}

\newcommand{\diff}[1][]{\mbox{d}#1}

\newcommand{\half}[1]{\ensuremath{\frac{#1}{2}}}
\newcommand{\intd}[1]{\int \!\! #1 \;}
\newcommand{\inv}[1]{\ensuremath{\frac{1}{#1}}}

\newcommand{\Sbindex}[1]{\ensuremath{\smallindex{\bar{\mathcal{S}}}{#1}}}
\newcommand{\Sbtext}[1]{\itindex{\bar{\mathcal{S}}}{#1}}
\newcommand{\Sindex}[1]{\ensuremath{\smallindex{\mathcal{S}}{#1}}}
\newcommand{\Stext}[1]{\itindex{\mathcal{S}}{#1}}

\newcommand{\derfrac}[2][]{\frac{\partial #1}{\partial #2}}

\newcommand{\itindex}[2]{\ensuremath{#1_{\mbox{\scriptsize{\itshape #2}}}}} 
\newcommand{\smallindex}[2]{\ensuremath{#1_{\scriptscriptstyle{#2}}}}

\newcommand{\varfrac}[2][]{\frac{\delta #1}{\delta #2}}

\DeclareMathOperator{\extdm}{d}
\newcommand{\extd}{\extdm \!}

\newcommand{\beqs}{\begin{equation*}}
\newcommand{\beq}{\begin{equation}}

\newcommand{\eeqs}{\end{equation*}}
\newcommand{\eeq}{\end{equation}}

\newcommand{\beqas}{\begin{eqnarray*}}
\newcommand{\beqa}{\begin{eqnarray}}

\newcommand{\eeqas}{\end{eqnarray*}}
\newcommand{\eeqa}{\end{eqnarray}}

\newcommand{\eq}[2]{\begin{equation} #1 \label{#2} \end{equation}}

\newcommand{\eps}{\varepsilon}
\newcommand{\al}{\alpha}
\newcommand{\be}{\beta}

\newcommand{\om}{\omega}

\newcommand{\blist}{\begin{itemize}}

\newcommand{\elist}{\end{itemize}}

\providecommand{\href}[2]{#2}

\newcommand{\twod}{$2D$}

\DeclareFontFamily{OT1}{rsfs}{}
\DeclareFontShape{OT1}{rsfs}{m}{n}{ <-7> rsfs5 <7-10> rsfs7 <10->rsfs10}{} 
\DeclareMathAlphabet{\mycal}{OT1}{rsfs}{m}{n}

\begin{document}
% TITLEPAGE
\renewcommand{\thefootnote}{\fnsymbol{footnote}}
\thispagestyle{empty}

\begin{titlepage}

\renewcommand{\thefootnote}{\fnsymbol{footnote}}

\hfill TUW--03--24

\hfill ESI 1381

%\hfill Vers. 2.0 -- DRAFT \today \\

\begin{center}
\vspace{0.5cm}

{\Large\bf Supersymmetric black holes in 2D dilaton\\ supergravity:
baldness and extremality}
\vspace{1.0cm}
% \vfill

{\bf L.\ Bergamin\footnotemark[1], D.\ Grumiller\footnotemark[2] and 
W. Kummer\footnotemark[3]
}
\vspace{7ex}

{Institut f\"ur
    Theoretische Physik \\ Technische Universit\"at Wien \\ Wiedner
    Hauptstr.  8--10, A-1040 Wien \\ Austria}
%% \vspace{2ex}
\vspace{1.5cm}

\footnotetext[1]{e-mail: {\tt bergamin@tph.tuwien.ac.at}}
\footnotetext[2]{e-mail: {\tt grumil@hep.itp.tuwien.ac.at}}
\footnotetext[3]{e-mail: {\tt wkummer@tph.tuwien.ac.at}}

\end{center}

\begin{abstract}

We present a systematic discussion of supersymmetric solutions of 2D
dilaton supergravity. In particular those solutions which retain
at least half of the supersymmetries are ground states with respect to the
bosonic Casimir function (essentially the ADM mass). Nevertheless, by tuning
the prepotential appropriately, black hole solutions may emerge with an arbitrary number of Killing
horizons. The absence of dilatino and gravitino hair is
proven. Moreover, the impossibility of supersymmetric dS ground states
and of nonextremal black holes is confirmed, even in the presence of a
dilaton. In these
derivations the knowledge of the general analytic solution of 2D
dilaton supergravity plays an important r\^{o}le. The latter result is addressed in the more general context
of gPSMs which have no supergravity interpretation.

Finally it is
demonstrated that the inclusion of non-minimally coupled matter, a step 
which is already nontrivial by itself, does not change these features in an
essential way.

\end{abstract}

%PACS numbers: ...

\vfill
\end{titlepage}

% BODY

\renewcommand{\thefootnote}{\arabic{footnote}}
\setcounter{footnote}{0}
\numberwithin{equation}{section}

\section{Introduction}

In the mid 1990s, during and after the ``second string revolution'', BPS
(Bogomolnyi-Prasad-Sommerfield \cite{Bogomolny:1976de}) black
holes (BHs) \cite{Gibbons:1981ja,Gibbons:1982fy} have attracted much interest because in particular they allow to
derive the BH entropy by counting D-brane microstates exploiting string dualities
(for reviews cf.\ e.g.\ \cite{Youm:1997hw}). We define a BPS
BH as a supergravity (SUGRA) solution respecting half of the supersymmetries
and exhibiting at least one Killing horizon in the bosonic line element.

A key reference for the properties of supersymmetric
solutions in 2D dilaton SUGRA is the work of Park and Strominger
\cite{Park:1993sd}. As shown by these authors for the SUGRA version of the
CGHS model \cite{Callan:1992rs}, as well as for a SUGRA extended generalized
2D dilaton theory, a certain vacuum solution can be defined with vanishing
fermions. A specific solution, which still retained one supersymmetry,
was constructed for the CGHS-related model. In the generic case the existence
of such a solution was proven, but it was not constructed.

Until quite recently a systematic study of {\em all} supersymmetric solutions
in 2D dilaton SUGRA theories has not been possible.
Recently two of the present authors have shown
\cite{Bergamin:2002ju,Bergamin:2003am} that the superfield formulation of
\cite{Park:1993sd} can be identified with the one of a certain subclass of
graded Poisson-Sigma models (gPSMs). General gPSMs are fermionic extensions of
bosonic PSMs which, in the present case, are taken to be 2D dilaton theories
of gravity \cite{Ertl:2000si}. This subclass of gPSMs has been dubbed minimal
field SUGRA (MFS) in ref.\ \cite{Bergamin:2003am}. The important consequence of this equivalence is that the known analytic solution in the MFS formulation \cite{Bergamin:2003am} represents the full solution for dilaton SUGRA \cite{Park:1993sd}, including all solutions with nonvanishing fermionic field components.

This permits us to attack in a systematic manner the problem of 2D SUGRA
solutions which retain at least one supersymmetry, the main goal of our present work. It
turns out that, although some information can be obtained from the symmetry
relations, the knowledge of the full solution is a necessary input. This is of
particular importance regarding the eventual existence of fermionic hair for
BHs\footnote{An early attempt to prove the validity of this no-hair conjecture
for the CGHS model can be found in \cite{Gamboa:1993xm}.}.

In doing this we are also able to incorporate SUGRA invariant matter, where a
special previous model \cite{Izquierdo:1998hg} is extended to generalized
dilaton theories and also transcribed into the more convenient MFS
formulation. Although no general analytic solution is possible when matter is
included, we show that there are no nonextremal BH solutions of 2D SUGRA
respecting half of the supersymmetries. This result is not unexpected:
``Pure'' SUGRA---i.e.\ not deformed by a (super-)dilaton field---does not
permit nonextremal supersymmetric BH solutions according to a simple but
elegant argument\footnote{A crucial ingredient is the continuation to the
  Euclidian domain and the observation that the absence of conical
  singularities enforces boundary conditions of thermal quantum field theory,
  which are not compatible with supersymmetry
  \cite{Girardello:1981vv} implying $T=0$ and
  hence extremality.} due to Gibbons \cite{Gibbons:1981ja}. Acutally, in 2D
\emph{dilaton} SUGRA a more direct proof is possible, which avoids the
continuation to Euclidian signature: First, we show that the body of the
Casimir function (a quantity in gPSM theories related to the energy of the
system) has to vanish and then we prove that these ground state solutions
cannot provide simple zeros of the Killing norm. {\em En passant} all
solutions respecting at least half of the supersymmetries are classified
including the ones of ref.\ \cite{Park:1993sd}.

The paper is organized as follows: After a short review of MFS models
(sect.\ \ref{sec:susy}) in sect.\ \ref{se:1} all ground states with
unbroken supersymmetry are discussed, which turn out to be constant dilaton
vacua. Such vacua appear nontrivially in some, but not all dilaton
theories. Sect.\ \ref{se:2} is devoted to a classification of solutions
respecting half of the supersymmetries. All of them imply a vanishing body for
the bosonic Casimir function. In sect.\ \ref{se:3} it is demonstrated that for
such ground states the Killing norm has to be positively semi-definite and
thus only extremal horizons may exist. This statement even can be extended to general gPSM theories. Sect.\ \ref{se:matter} discusses the
coupling of MFS to matter degrees of freedom and sect.\ \ref{se:mattergs}
generalizes the result of sect.\ \ref{se:3} to the one including matter.

The
notation is explained in app.\ \ref{sec:notation}. 
Equations of motion and some aspects of the exact solution are contained in
app.\ \ref{sec:eom}. The subject of app.\  \ref{sec:integrals} are some key
formulas \cite{Bergamin:2003am} relating MFS to the superfield formulation of 2D
dilaton SUGRA \cite{Park:1993sd}, which are needed for the construction of the matter couplings.
\section{Minimal field SUGRA (MFS)}\label{sec:susy}
General 2D dilaton SUGRA can be formulated in terms of a gPSM
\cite{Ikeda:1993aj,Ertl:2000si}.
Its action with target space variables $X^I$, gauge fields $A_I$ and Poisson tensor $P^{IJ}$
\begin{equation}
  \label{eq:gPSMaction}
    \Stext{gPSM} = \int_M \extd X^I \wedge A_I + \half{1} P^{IJ} A_J \wedge
    A_I 
\end{equation}
is invariant under the symmetry transformations
\begin{align}
\label{eq:symtrans}
  \delta X^{I} &= P^{IJ} \ve _{J}\ , & \delta A_{I} &= -\mbox{d} \ve
  _{I}-\left( \partial _{I}P^{JK}\right) \ve _{K}\, A_{J}
\end{align}
as a consequence of the graded non-linear Jacobi identity
\begin{equation}
  \label{eq:ref5}
    P^{IL}\partial _{L}P^{JK}+ \mbox{\it 
g-perm}\left( IJK\right) = 0\ .
\end{equation}
Not every gPSM may be used to describe 2D SUGRA. A subclass of
appropriate models
(called minimal field SUGRA, MFS) has been identified in
\cite{Bergamin:2002ju, Bergamin:2003am} for $N=(1,1)$ SUGRA. It contains a
dilatino $\chi_\alpha$ and a gravitino $\psi_{\alpha}$, both of which are
Majorana spinors ($X^I = (\phi, X^a, \chi^\alpha)$,
$A_I = (\omega, e_a, \psi_\alpha)$, $Y = X^{++} X^{--}$):
\begin{gather}
\label{eq:ll}
\begin{alignat}{2}
  P^{a \phi} = X^b {\epsilon_b}^a &\qquad  P^{\alpha \phi} &= - \half{1}
  \chi^\beta {\gthree_\beta}^\alpha
\end{alignat}\medsp
\label{eq:mostgensup}
  P^{ab} = \biggl( V + Y Z - \half{1} \chi^2 \Bigl( \frac{VZ + V'}{2u} +
  \frac{2 V^2}{u^3} \Bigr) \biggr) \epsilon^{ab} \medsp
  P^{\alpha b} = \frac{Z}{4} X^a
    {(\chi \gamma_a \gamma^b \gthree)}^\alpha + \frac{i V}{u}
  (\chi \gamma^b)^\alpha\medsp
\label{eq:mostgensuplast}
  P^{\alpha \beta} = -2 i X^c \gamma_c^{\alpha \beta} + \bigl( u +
  \frac{Z}{8} \chi^2 \bigr) \gthree^{\alpha \beta}
\end{gather}
$V$, $Z$ and the prepotential $u$ are functions of the dilaton field $\phi$
and obey the relation
\begin{equation}
V=-\frac 18 \left[(u^2)'+u^2Z\right]\ . \label{eq:sol1}
\end{equation}
This Poisson tensor leads to the action
\begin{multline}
  \label{eq:mostgenaction}
  \Stext{MFS} = \int_{\mathcal{M}} \bigl( \phi \diff \omega + X^a D e_a + \chi^\alpha D
  \psi_\alpha + \epsilon \biggl( V + Y Z - \half{1} \chi^2 \Bigl( \frac{VZ + V'}{2u} +
  \frac{2 V^2}{u^3} \Bigr) \biggr) \medsp
   + \frac{Z}{4} X^a
    (\chi \gamma_a \gamma^b e_b \gthree \psi) + \frac{i V}{u}
  (\chi \gamma^a e_a \psi) \medsp
   + i X^a (\psi \gamma_a \psi) - \half{1} \bigl( u +
  \frac{Z}{8} \chi^2 \bigr) (\psi \gthree \psi) \bigr)\ .
\end{multline}
An important class of simplified models is described by the special choice $\bar{Z}
= 0$. Following the nomenclature of \cite{Bergamin:2003am} it is
called MFS$_0$ and barred variables are used.
As can be verified easily the action \eqref{eq:mostgenaction} and
$\Stext{MFS$_0$}$ are related by a
conformal transformation of the fields ($Q'(\phi) = Z(\phi)$)
\begin{align}
\label{eq:ct1}
  \phi &= \bar{\phi}\ , & X^a &= e^{- \half{1} Q(\phi) } \bar{X}^a\ , & \chi^\alpha
  &= e^{- \inv{4} Q(\phi)} \bar{\chi}^\alpha\ , \medsp
\label{eq:ct2}
  \omega &= \bar{\omega} + \half{Z} \bigl(\bar{X}^b \bar{e}_b + \half{1} \bar{\chi}^\beta \bar{\psi}_\beta
  \bigr)\ , & e_a &= e^{\half{1} Q(\phi)} \bar{e}_a\ , & \psi_\alpha &= e^{
  \inv{4} Q(\phi)} \bar{\psi}_\alpha\ .
\end{align}
The symmetry transformations \eqref{eq:symtrans} for fermionic $\epsilon$ with
\eqref{eq:ll}-\eqref{eq:mostgensuplast} become:
\begin{align}
  \label{eq:gPSMtransf}
\delta \phi &= \half{1} (\chi \gthree \ve) \, \medsp
\delta X^a &= - \frac{Z}{4} X^b (\chi \gamma_b \gamma^a \gthree \ve) - \frac{i
  V}{u} (\chi \gamma^a \ve) \label{eq:gPSMtransf2}\, \medsp
\delta \chi^\alpha &= 2i X^c (\ve \gamma_c)^\alpha - \bigl(u + \frac{Z}{8}
\chi^2 \bigr) (\ve \gthree)^\alpha \label{eq:gPSMtransf3}\, \medsp
\delta \omega &= \frac{Z'}{4} X^b (\chi \gamma_b \gamma^a \gthree \ve) e_a + i
\bigl(\frac{V}{u}\bigr)' (\chi \gamma^a \ve) e_a + \bigl(u' + \frac{Z'}{8}
\chi^2 \bigr) (\ve \gthree \psi) \label{eq:gPSMtransf4}\, \medsp
\delta e_a &= \frac{Z}{4} (\chi \gamma_a \gamma^b \gthree \ve) e_b - 2i (\ve
\gamma_a \psi) \label{eq:gPSMtransf5}\, \medsp
\delta \psi_\alpha &= - (D \ve)_\alpha + \frac{Z}{4} X^a (\gamma_a \gamma^b
\gthree \ve)_\alpha e_b +  \frac{i
  V}{u} (\gamma^b \ve)_\alpha e_b + \frac{Z}{4} \chi_\alpha (\ve \gthree
\psi)\label{eq:gPSMtransflast}
\end{align}
By eliminating $X^a$ and the torsion dependent part of the spin connection a
new action (MFDS) in terms of dilaton, dilatino, zweibein and gravitino is
obtained\footnote{\label{fn:tildedef}Quantities with a tilde refer to the
  dependent spin connection $\tilde{\omega}_a = \epsilon^{mn} \partial_n e_{ma} - i \epsilon^{mn} (\psi_n
  \gamma_a \psi_m)$, i.e.\ $\tilde{\sigma}_\alpha = \ast
    (\tilde{D} \psi)_\alpha$ and $\tilde{R} = 2 \ast \extd
    \tilde{\omega}$. The quantity $\tilde{\sigma}_\alpha$
is the fermionic partner of the curvature scalar $\tilde{R}$.} 
\begin{multline}
\label{eq:mostgenaction2}
  \Stext{MFDS} = \intd{\diff{^2 x}} e \biggl( \half{1} \tilde{R} \phi +
  (\chi \tilde{\sigma}) + V - \inv{4 u} \chi^2 \bigl( V Z + V' + 4
  \frac{V^2}{u^2} \bigr) \medsp
  - \half{1} Z \Bigl( \partial^m \phi \partial_m \phi + \half{1} (\chi \gthree
  \psi^m) \partial_m \phi + \half{1} \epsilon^{mn} \partial_n \phi (\chi
  \psi_m)\Bigr) \medsp
  - \frac{i V}{u} \epsilon^{mn} (\chi \gamma_n \psi_m) + \half{u}
  \epsilon^{mn} (\psi_n \gthree \psi_m) \biggr)\ .
\end{multline}
Its supersymmetry transformations read
\begin{align}
  \delta \phi &= \half{1} (\chi \gthree \ve)\ , \label{eq:elimtransf}\medsp
  \delta \chi^\alpha &= - 2i \epsilon^{mn} \bigl( \partial_n \phi + \half{1}
  (\chi \gthree \psi_n) \bigr)  (\ve \gamma_m)^\alpha - \bigl(u + \frac{Z}{8}
\chi^2 \bigr) (\ve \gthree)^\alpha\ , \label{eq:elimtransf2}\medsp
  \delta {e_m}^a &= \frac{Z}{4} (\chi \gamma^a \gamma^b \gthree \ve) e_{mb} - 2i (\ve
\gamma^a \psi_m)\ , \label{eq:elimtransf3}\medsp
  \delta \psi_{m \alpha} &= - (\tilde{D} \ve)_\alpha + \frac{iV}{u} (\gamma_m
  \ve)_\alpha + \frac{Z}{4} \bigl( \partial^n \phi (\gamma_m \gamma_n
  \ve)_\alpha + \half{1} (\psi_m \gamma^n
  \chi) (\gamma_n \gthree \ve)_\alpha \bigr)\ .
\label{eq:elimtransflast}
\end{align}
The action \eqref{eq:mostgenaction2} has been shown to be equivalent \cite{Bergamin:2003am} to the
general dilaton superfield SUGRA of Park and Strominger \cite{Park:1993sd}
(cf.\ app.\ \ref{sec:integrals}).
\section{Both supersymmetries unbroken}\label{se:1}

From the MFS supersymmetry transformations in sect.\ \ref{sec:susy}
one can read off different conditions for solutions respecting full 
supersymmetry.
From \eqref{eq:gPSMtransf} and \eqref{eq:gPSMtransf5} follows\footnote{Conventions and light cone coordinates are summarized in app.\ \ref{sec:notation}.} $\chi^+ = \chi^-
= \psi_+ = \psi_- = 0$. The two terms in \eqref{eq:gPSMtransf3} are linearly
independent and thus $X^{++} = X^{--} = u = 0$ as
well. In addition \eqref{eq:gPSMtransflast} implies that the transformation
parameters must be covariantly constant: $ (D \ve)_\alpha = 0$. For a solution
where both supersymmetries are unbroken the Poisson tensor
\eqref{eq:ll}-\eqref{eq:mostgensuplast} vanishes
identically.

The equations of motion imply that the dilaton $\phi$ has to be a constant. It is restricted to a solution of the equation $u=0$. Such constant dilaton vacua (CDV) are, for instance, encountered \cite{Grumiller:2003ad} in the ``kink'' solution of the dimensionally reduced gravitational Chern-Simons term \cite{Guralnik:2003we}.

We recall in app.\ \ref{sec:eom} that a key ingredient of the solution is the
conserved Casimir function. At this point its additive ambiguity can be fixed:
supersymmetry covariance requires that solutions respecting both supersymmetries have vanishing Casimir function. This means that eventual additive constants in \eqref{sol:14},\eqref{sol:12} are absent.

Positivity of energy would imply $u^2\geq 8Y$ because the Casimir function $C_B$ is
related to the negative ADM mass (cf.\ section 5 of ref.\ \cite{Grumiller:2002nm}). If this inequality is saturated the ground
state is obtained. Eq.\ \eqref{sol:14} in particular implies that all CDV solutions
with $u = 0$ have vanishing body of the Casimir function.\footnote{CDV 
solutions with $u \neq 0$ must obey $u' / u = - \half{1} Z$, which leads to 
$C_S = 0 $ while $C_B \neq 0$. Clearly, they cannot respect both 
supersymmetries.}

As an illustration we consider a two parameter family of models (the so-called
``$ab$-family'') encompassing most of the relevant ones
\cite{Katanaev:1997ni}. Among other solutions, BHs immersed in Minkowski,
Rindler or (A)dS space can be described. This family is defined by (\ref{eq:mostgenaction}) with 
\eq{
Z(\phi)=-\frac a\phi\,,\quad u(\phi)=c\phi^{\al}\,, \quad \al,a,c\in\mathbb{R}\,.
}{eq:sol3}
Supersymmetry restricts the constant $B = c^2 (b+1)/4$ in the potential (\ref{eq:sol1})
\eq{
V(\phi)=-\frac B2 \phi^{a+b}\,,\quad \al=\frac{a+b+1}{2}
}{sol:4}
to $B>0$ if $b>-1$ and to $B<0$ if $b<-1$. The curvature scalar of the ground state geometry is given by
\eq{
R=\frac{bc^2}{2}(\al-a)X^{2(\al-1)}\,.
}{sol:5}
The Minkowski Ground State (MGS) condition\footnote{It means simply that for
  vanishing bosonic Casimir function the bosonic line element is diffeomorphic
  to the one of Minkowski space.} reads $\al=a$, Rindler space follows for $b=0$ and (A)dS means $\al=1$. In the latter case supersymmetry restricts the curvature scalar $R=c^2(1-a)^2/2$ to positive values, which in our notation implies AdS.

For fully supersymmetric solutions of the $ab$-family the only possible values for the dilaton are $\phi=0$ or $|\phi|=\infty$ (depending on the value of $\al$), unless $c=0$, in which case the prepotential $u$ vanishes identically.

\section{One supersymmetry unbroken}\label{se:2}

\subsection{Casimir function $\boldsymbol{C_B=0}$}

The symmetry transformation $\delta \phi = 0$ of the dilaton from (\ref{eq:gPSMtransf}) implies 
\eq{
\chi^+\ve_+=\chi^-\ve_-\,.
}{sol:13}
The vanishing of \eqref{eq:gPSMtransf3} and \eqref{eq:gPSMtransf5}
leads to
\begin{gather}
\label{eq:epscond1}
\begin{alignat}{2}
  u\ve_- &= - 2 \sqrt{2} X^{++} \ve_+\ , &\qquad u\ve_+ &= - 2 \sqrt{2} X^{--}
  \ve_-\ ,
\end{alignat}\medsp
\label{eq:epscond3}
\ve \gamma_a \psi = -i \frac{Z}{8} {\epsilon_a}^b e_b (\chi \ve)\ .
\end{gather}
In \eqref{eq:epscond1} terms proportional to $\chi^2\eps$ have to vanish as a
consequence of \eqref{sol:13}.
Eqs.\ \eqref{eq:epscond1} require
\begin{equation}
  \label{eq:epscond2}
  Y = \inv{8} u^2\ ,
\end{equation}
which in turn implies that the body of the Casimir  function (\ref{sol:14})
vanishes. In this sense, BPS like states are always ground states. 
Notice that eq.\ \eqref{eq:epscond2} remains
valid in the case $u = 0$, implying that at least one component of
$X^a$ vanishes as well and vice versa.

It is worthwhile emphasizing that \eqref{eq:epscond2} corresponds to a
vanishing determinant
\begin{equation}
  \label{eq:ref1}
  \Delta = \det 
  \begin{pmatrix}
    -2 \sqrt{2} X^{++} & -u \\ -u & -2 \sqrt{2} X^{--} 
  \end{pmatrix}
\end{equation}
of the bosonic part of \eqref{eq:mostgensuplast} (cf.\
\cite{Ertl:2000si}). $\Delta = 0$ must hold for any solution that respects at
least one supersymmetry.

\subsection{Classification of solutions}

\subsubsection{Vanishing fermions}

We first consider the case of vanishing dilatino $\chi^\alpha$ and gravitino $\psi_\alpha$. Then all three quantities $u,X^a$ have to be nonvanishing or otherwise the solution with full supersymmetry of sect.\ \ref{se:1} is recovered.
With the conditions (\ref{eq:epscond1}) and hence also (\ref{eq:epscond2}) the variations (\ref{eq:gPSMtransf})-(\ref{eq:gPSMtransf5}) vanish, while \eqref{eq:gPSMtransflast} equal zero
represents two differential equations for $\ve_+$, resp.\ $\ve_-$. By inserting
the explicit solution (sect.\ 6 of ref.\ \cite{Bergamin:2003am}) it can be checked
that both are identical. A straightforward
calculation, without using any further restrictions on the different
variables involved, yields
\begin{equation}
  \label{eq:epsdg1}
  \extd \ve_+ + \left( \frac{\extd X^{++}}{2 X^{++}} - (\frac{u'}{u} + \half{1}
  Z) \extd \phi \right) \ve_+ = 0\ ,
\end{equation}
possessing the general solution ($Q$ is defined in (\ref{eq:sol2}))
\begin{equation}
  \label{eq:epssol1}
  \ve_+ = e^{\half{1} Q} \frac{u}{\sqrt{X^{++}}} \tilde{\ve} \ .
\end{equation}
Here $\tilde{\ve}$ is a spinorial integration constant. $\ve_-$ is obtained
via \eqref{eq:epscond1}. 
Thus, all solutions without fermion fields, exhibiting one supersymmetry, 
leave a linear combination of the two supersymmetries unbroken,
in agreement with the discussion in ref.\ \cite{Park:1993sd}.

There exists one special case of \eqref{eq:epsdg1} where an even simpler
solution exists: If $Z = 0$ (MFS$_0$ in the parlance of
ref.\ \cite{Bergamin:2003am}) and if in addition $u$ is a constant, the
differential equation \eqref{eq:epsdg1} simply says that the symmetry
parameter is covariantly constant. This model is generalized teleparallel dilaton gravity. As $u$ is simply a cosmological constant, it drops
out in the constraint algebra, and therefore this case has been referred to
as rigid supersymmetry in ref.\ \cite{Ertl:2000si}.

\subsubsection{Nonvanishing fermions and no-hair theorem}

Further supersymmetric solutions are possible for nonvanishing fermions, a
situation not considered in ref.\ \cite{Park:1993sd}, but relevant for the
question of fermionic hair\footnote{No-hair theorems are a recurrent theme in
  BH physics (ref.\ \cite{Frolov:1998} and refs.\ therein).}.

Assuming
that at least one component of $\chi_\alpha$ is different from zero,
e.g. $\chi^+ \neq 0$, eq.\ \eqref{sol:13} can have the chiral solution $\ve_+ =
\chi^- = 0$, while $\ve_- \neq 0$ provides the remaining supersymmetry, or it can
relate $\ve_+$ to $\ve_-$ by means of $\chi^+ \neq 0$ and $\chi^- \neq
0$. 

Solutions of the first type are almost trivial, because they require $u =
X^{--} = 0$. Since the dilaton must not be constant (or else the CDV case
would be recovered) this implies that $u$ must vanish identically, and not
just at a certain value of $\phi$. All fermions must be of one chirality, in particular
$\eps_+=\chi^-=\psi_-=0$. The quantities $X^{++}, e_{++}, \om, \chi^+$ must be
covariantly constant and may contain soul contributions. The dilaton has to
fulfill the linear dilaton vacuum condition $\extd\phi=\rm const.$ Only $e_{--}$
and $\eps_-$ are slightly nontrivial and can be deduced from
\begin{align}
\label{sol:15}
(De)_{--} &= Z \mu e_{--}\ , & (D\eps)_- &= \frac Z2 \mu \eps_-\ , & \mu &= (X^{++}e_{++}+\half{1} \chi^+\psi_+)\ .
\end{align}
The bosonic line element is flat and obviously the Casimir function is identically zero.

For the remaining class of solutions with {\em both} components of $\chi$ and of $\ve$
different from zero the body of the Casimir function still vanishes, but the soul can be
non-vanishing as from \eqref{eq:gPSMtransf2} and \eqref{sol:12}:
\begin{align}
  \label{eq:epscond4}
  Z &= - \frac{u'}{u} & C_S &= \inv{32} e^Q u' \chi^2
\end{align}
This is equivalent to the MGS condition mentioned in section \ref{se:1} below (\ref{sol:5}). 
Thus, only MGS models are allowed and since the solution has to
be the ground state, the bosonic part of the geometry is trivially Minkowski space. 

Therefore, {\em a solution with nonvanishing fermions must have a trivial
  bosonic background} (this feature is true also for the first
  type). Consequently there exist no BPS BHs with fermionic hair.

There remains a technical subtlety about the nonchiral solutions. As both
components of $\ve$ are non-zero, \eqref{eq:epscond1} implies that all
interesting cases have $u$, $X^{++}$ and $X^{--}$ different from zero. Actually, the solutions presented in \cite{Bergamin:2003am} do not cover this case, because the one for $C \neq 0$ ((6.9)-(6.13) of
\cite{Bergamin:2003am}), depending on $C^{-1}$, cannot be used in the present case, as the inverse of
a pure soul is ill-defined. The solution for $C = 0$ ((6.17)-(6.21) of
\cite{Bergamin:2003am}) depends on an arbitrary fermionic gauge
potential $\tilde{A} = -\extd f$, which is the gauge potential associated with
the additional fermionic Casimir function (eq.\ (6.15) of
\cite{Bergamin:2003am}) appearing in that case. Analyzing the e.o.m.-s (app.\
\ref{sec:eom} and cf.\ also \cite{Ertl:2000si}) for $C_B = 0$ but $C_S
\neq 0$, this fermionic Casimir function does no longer appear, but the solution remains
valid if the gauge potential $\tilde{A} = -\extd f$ obeys the constraint
$\chi^2 \tilde{A} = 0$.

Finally \eqref{eq:gPSMtransflast} again leads to the differential equation
\begin{equation}
  \label{eq:epsdg2}
  D \ve_+ - \left(\frac{Z}{16} \frac{u^2}{X^{++}} e_{--} - \half Z X^{++}
  e_{++}\right) \ve_+ = 0\ .
\end{equation}
Inserting the solution described above all trilinear and higher spinorial terms are found to vanish. Thus, the solution of (\ref{eq:epsdg2}) reduces to (\ref{eq:epssol1}).

\section{No nonextremal BPS black holes}\label{se:3}

\subsection{MFS models}\label{se:3.1}

It has been shown in section \ref{se:2} that the body of the bosonic Casimir function has to vanish for all solutions respecting at least half of the supersymmetries. We will focus first on BPS solutions where either $X^{++}\neq 0$ or $X^{--}\neq 0$ (or both). 

In the bosonic case ground state solutions $C=0$ imply for the line element (cf.\ e.g.\ eq.\ (3.26), (3.27) of \cite{Grumiller:2002nm}, $\otimes$ denotes the symmetrized tensor product)
\eq{
(\extd s)^2=2\extd F\otimes (\extd r-e^{Q(\phi)}w(\phi)dF)\,,\quad \extd r=e^{Q(\phi)}\extd\phi\,.
}{sol:10}
Obviously, there exists always a Killing vector $\xi^\mu\partial_\mu=\partial_F$ the norm of which is given by $K=-2e^{Q(\phi)}w(\phi)$.
By choosing the function $w(\phi)$ appropriately, nonextremal Killing horizons are possible, e.g.\ for $Q=0, -2w=1-2m/\phi$ a Schwarzschild-like BH emerges. 

In SUGRA, however, (\ref{eq:sol2}) implies a negative (semi-)definite $w$ and hence the Killing norm $K$ can only have zeros of even degree:
\eq{
K(\phi)=-2e^{Q(\phi)}w(\phi) = \left(\frac 12 u(\phi)e^{Q(\phi)}\right)^2\geq 0
}{sol:17}
Thus, if a Killing horizon exists it has to be an extremal one, which confirms
the general proof using thermal field theory arguments \cite{Gibbons:1981ja}. For instance, with
$u=2(\sqrt{\phi}-M), Z(\phi)=-1/(2\phi)$ the line element reads $(\extd
s)^2=2\extd F\otimes(\extd r+(1-M/r)^2\extd F)$, which is the two-dimensional
part of an extremal Reissner-Nordstr\"om BH\footnote{In dilaton (super)gravity
  the number and types of horizons can be
adjusted by selecting a certain behavior of the functions $w$ and (to a
lesser extent) $Q$, which enter the Killing norm \eqref{sol:17}. In many cases of
physical relevance extremality is induced by tuning of certain
charges/constants of motion, but we emphasize that the explicit presence
of additional (gauge) fields by no means is necessary for extremality. For
instance, the  Reissner-Nordstr\"om BH can be constructed either from spherically reduced
Einstein-Maxwell theory by tuning the two Casimir functions (mass and
charge) accordingly, but it is also possible to provide an effective
description where one of these constants, the charge, enters as a
parameter of the action rather than a constant of motion.
}.

One can trivially generalize this result to all ground state solutions which are not CDV, i.e.\ non-supersymmetric solutions of {\twod} dilaton SUGRA with vanishing body of the Casimir function and non-constant dilaton, because the key inequality (\ref{sol:17}) still holds.

Finally, the simpler CDV case $X^{++}=0=X^{--}$ shall be addressed. As shown
in the previous two sections CDV implies that both supersymmetries are either
broken or unbroken. The equations of motion imply a vanishing body of torsion
and a constant body of curvature. Thus -- as in the bosonic case (cf.\ sect.\
2.1 in \cite{Grumiller:2003ad}) -- only (A)dS, Rindler\footnote{If curvature
  is non-vanishing the Rindler term can always be absorbed by a linear
  redefinition of the coordinates $r\to r' = \al r +\be$, $F \to F'=F/\al$,
  $\al\neq 0$. For vanishing curvature the Minkowski term can always be
  absorbed by a similar redefinition.} or Minkowski spaces are
possible. However, supersymmetry provides again an obstruction: curvature is
proportional to $(u')^2$ and thus again the dS case, together with the
possibility of nonextremal Killing horizons, is ruled out by
supersymmetry.\footnote{In fact, there is one very trivial possibility that
  remains for a CDV solution with vanishing curvature which allows for the
  existence of exactly one nonextremal Killing horizon: the line element
  $(\extd s)^2=2\extd F\otimes(\extd r+b r \extd F)$, $b\neq 0$ contains a
  nonextremal (Rindler) horizon at $r=0$. However, this is neither a BPS state
  nor a true BH solution.}
\subsection{Generic gPSM gravity}\label{se:3.2}
The question of nonextremal BPS BHs may be addressed in a more general context,
namely for gPSM gravity that does not belong to the MFS class. Here we
consider generic Poisson tensors with local Lorentz invariance implemented as in eq.\
\eqref{eq:ll}. In addition the fermionic extension $P^{\alpha \beta}$ must have full rank almost
everywhere in the space of solutions with the notable exception of those which still obey eq.\ \eqref{sol:13} and where consequently the determinant $\Delta$ (analogous to \eqref{eq:ref1}) vanishes. These solutions will be called BPS because they still respect half of the fermionic symmetries.

In a generic gPSM \eqref{eq:mostgensuplast} is replaced by $P^{\alpha \beta} =
v^{\alpha \beta} + \chi^- \chi^+ v_2^{\alpha \beta}$ with
\begin{equation}
  \label{eq:ref2}
  v^{\alpha \beta} = 
  \begin{pmatrix}
    \sqrt{2} X^{++} (\tilde{u} - \hat{u}) & -u \\ -u & \sqrt{2} X^{--}
    (\tilde{u} + \hat{u})
  \end{pmatrix}\ ,
\end{equation}
where $u$, $\tilde{u}$ and $\hat{u}$ are functions of $\phi$ and $Y$
\cite{Ertl:2000si}. $v_2^{\alpha \beta}$ is determined by the Jacobi
identity \eqref{eq:ref5}. Also in the bosonic potential
\begin{equation}
  \label{eq:ref4}
  P^{ab} = \epsilon^{ab} \bigl( v(\phi, Y) + \chi^- \chi^+ v_2(\phi, Y) \bigr)
\end{equation}
the body is an independent function, while $v_2$ again follows from the Jacobi
identity. Vanishing determinant of \eqref{eq:ref2} implies
\begin{equation}
  \label{eq:ref3}
  Y = \frac{u^2}{2(\tilde{u}^2 - \hat{u}^2)}\ .
\end{equation}
On the other hand, the Killing norm is proportional to $Y$ (cf.\ eq.\ (36) in \cite{Klosch:1996fi}) and thus
nonextremal horizons are possible if $\tilde{u}$ and $\hat{u}$ are both
non-zero and field-dependent\footnote{These states in general are not
  ground-states in the sense of sect.\ \ref{se:1}.} (e.g. $u = a+b \phi$,
$\tilde{u} + \hat{u} = a + b \phi$, $\tilde{u} - \hat{u} = c$; $a,b,c\in\mathbb{R}$). However, following the arguments of ref.\ \cite{Gibbons:1981ja} all BPS BHs
should be extremal in generalized gPSM gravity theories as well.\footnote{The constraints from gPSM
symmetries are first class and free of ordering problems
\cite{Grosse:1992vc,Bergamin:2002ju,Bergamin:2001}. Therefore, on the
constraint surface the unbroken
fermionic symmetry still commutes with the Hamiltonian, which is the central
ingredient in the argument by Gibbons.} 

This apparent contradiction is resolved by
investigating singularity obstructions on the Poisson tensor (cf.\ sect.\ 3 in
ref.\ \cite{Ertl:2000si} and ref.\ \cite{Bergamin:2002ju}). Solving the Jacobi
identity
\eqref{eq:ref5} with $u$,
$\tilde{u}$, $\hat{u}$ and $v$ as a given input, all remaining functions $P^{a
\beta}$, $v_2$ in \eqref{eq:ref4} and $v^{\alpha \beta}_2$
are proportional to $\Delta^{-1}$. Only for very special relations among the four
free functions the inverse powers of $\Delta$ can be removed. It turns out that these relations imply extremality of eventual horizons appearing in BPS solutions. Consequently, ``BPS states'' of generalized gPSM gravity theories with nonextremal horizons are {\em singular}
solutions of the equations of motion\footnote{Similar states with
  singularities in the gravitino sector at the horizon had been found in 4D supergravity as well,
  cf.\ the discussion in ref.\ \cite{Gibbons:1981ja}.}. 

\section{Extension with conformal matter}
\label{se:matter}
We will prove in the next section that the conclusions of the sect.\ \ref{se:3.1} do not change
when conformal matter is coupled to the dilaton SUGRA system. As these
conclusions rely on the details of the symmetry transformations and the
conserved quantities, in a first step the extension to MFS
with matter fields is introduced in this section. To this end the close relation between MFS and the models obtained from superspace
\cite{Bergamin:2003am} is used.  In
superspace non-minimally coupled conformal matter is described by the Lagrangian   
\begin{equation}
  \label{eq:cmsup}
  \Sindex{(m)} = \inv{4} \intd{\diff{^2 x} \diff{^2 \theta}} E P(\Phi) D^\alpha M
  D_\alpha M\ .
\end{equation}
Here $P(\Phi)$ is a function of the dilaton superfield $\Phi$ (cf.\
\eqref{eq:dilatonsf}) and for the $\theta$-expansion of the matter multiplet
$M$ we write\footnote{Whenever the distinction between superfield components and
  MFS fields is important, underlined symbols are used for the former
  ones. However, for simplicity this is omitted in most formulas of this
  section, as the matter action is invariant under the redefinition
  \eqref{eq:genequiv}, while all other identifications are trivial.}
\begin{equation}
  M = f - i \theta \lambda + \half{1} \theta^2 H\ .
\end{equation}
Integrating out superspace one arrives at (cf.\ app.\ \ref{sec:integrals})
\begin{equation}
  \label{eq:cmsup2}
  \begin{split}
    \Sindex{(m)} &= \intd{\diff{^2 x}} e \Biggl[ P(\phi) \Bigl(\half{1}
    (\partial^m f \partial_m f + i \lambda \gamma^m \partial_m \lambda + H^2)
    \medsp
    &\quad + i (\psi_n \gamma^m \gamma^n \lambda) \partial_m f + \inv{4} (\psi^n
    \gamma_m \gamma_n \psi^m) \lambda^2 \Bigr) \medsp
    &\quad + \inv{4} P'(\phi) \Bigl( i (\lambda \gthree \chi) H -
    (\chi \gthree \gamma^m \lambda) \partial_m f - F \lambda^2
    \Bigr)\medsp
    &\quad
    - \inv{32} P''(\phi) \chi^2 \lambda^2 \Biggr]\ .
  \end{split}
\end{equation}
The action \eqref{eq:cmsup2} depends on the auxiliary fields $H$ from the
matter multiplet and $F$ from the dilaton multiplet. $H$ can be integrated out
without detailed knowledge of the geometric part of the action. To integrate
out $F$, however, $u(\Phi)$ and $Z(\Phi)$ in \eqref{eq:ps1} must be
specified\footnote{As $F$ appears in a term $\propto P'$ the restriction to $\bar{Z}
= 0$ in the following is not necessary when considering minimally coupled
matter.}.
\subsection{Matter extension at $\mathbf{\bar{Z} = 0}$\label{sec:5.1}}
 A particulary simple situation is realized by choosing $\bar{Z} = 0$
(following the notation of \cite{Bergamin:2003am}, barred variables are used
for this special case throughout). Then
the action \eqref{eq:ps1} is bilinear in $\bar{A}$ and $\bar{F}$ and the elimination
condition \eqref{eq:psauxfield} is modified according to $\bar{A} = -
\bar{u}'/2 + \bar{P}' \bar{\lambda}^2 / 4$, $\bar{F} = - \bar{u}/2$. 
The part of the action independent of the matter field retains the form
\eqref{eq:ps3} with $\bar{Z} = 0$,  while \eqref{eq:cmsup2} after elimination of all auxiliary fields
becomes:
\begin{equation}
  \label{eq:cmosh}
  \begin{split}
    \Sbindex{(m)} &= \intd{\diff{^2 x}} \bar{e} \Biggl[ \bar{P} \Bigl(\half{1}
    (\partial^m \bar{f} \partial_m \bar{f} + i \bar{\lambda} \gamma^m \partial_m \bar{\lambda})
    + i (\bar{\psi}_n \gamma^m \gamma^n \bar{\lambda}) \partial_m \bar{f}\medsp
    &\quad  + \inv{4} (\bar{\psi}^n
    \gamma_m \gamma_n \bar{\psi}^m) \bar{\lambda}^2 \Bigr)  +
    \frac{\bar{u}}{8} \bar{P}' \bar{\lambda}^2 - \inv{4} \bar{P}' 
    (\bar{\chi} \gthree \gamma^m \bar{\lambda}) \partial_m \bar{f} \medsp
    &\quad
    - \inv{32} \Bigl(\bar{P}''  - \inv{2} \frac{\bigl[\bar{P}' \bigr]^2}{\bar{P}}\Bigr) \bar{\chi}^2 \bar{\lambda}^2 \Biggr]
  \end{split}
\end{equation}
The symmetry transformations of the matter fields $\bar{f}$ and
$\bar{\lambda}_\alpha$ after elimination of $\bar{H}$ read (it should be noted
\cite{Bergamin:2003am} that the symmetry parameters $\ve$ and $\tve$ are
different in general)
\begin{align}
\label{eq:cmtransf}
  \bar{\delta} \bar{f} &= i (\bar{\tve} \bar{\lambda})\ , &
  \bar{\delta} \bar{\lambda}_\alpha &=  \bigl( \partial_m \bar{f} + i
  (\bar{\psi}_m \bar{\lambda}) \bigr) (\gamma^m \tve)_\alpha - \inv{4} \frac{P'}{P}
  (\bar{\lambda} \gthree \bar{\chi}) \bar{\tve}_\alpha\ ,
\end{align}
while the zweibein, the dilaton and the dilatino still transform according to
\eqref{eq:psostransf}, \eqref{eq:psostransf3} and
\eqref{eq:psostransflast} resp. The transformation rule for the gravitino changes,
as it depends on the auxiliary field $\bar{A}$:
\begin{equation}
  \label{eq:psicmtr}
    \delta {\underline{\bar{\psi}}_m}^\alpha = - (\Tilde{\Bar{D}} \bar{\tve})^\alpha +
    \frac{i}{4} \bigl( \bar{u}' - \half{1} \bar{P}'  \bar{\lambda}^2 \bigr)  
    (\bar{\tve} \gamma_m)^\alpha
\end{equation}

When working with the MFS formulation of the geometric part, it is
advantageous to formulate the matter action \eqref{eq:cmosh} in terms of differential forms as
well:
\begin{equation}
  \label{eq:cmdiffform}
  \begin{split}
    \Sbindex{(m)} &= \int_M \biggl[ \bar{P}  \Bigl( \half{1} \extd \bar{f} \wedge
    \ast \extd \bar{f} + \half{i}
    \bar{\lambda} \gamma_a \bar{e}^a \wedge \ast \extd \bar{\lambda} + i \ast(\bar{e}_a \wedge \ast \extd \bar{f}) \bar{e}_b \wedge \ast \bar{\psi} \gamma^a
    \gamma^b \bar{\lambda} \medsp
    &\quad + \inv{4} \ast (\bar{e}_b \wedge \ast \bar{\psi}) \gamma^a \gamma^b
    \bar{e}_a \wedge \ast \bar{\psi} \bar{\lambda}^2 \Bigr) +
    \frac{\bar{u}}{8} \bar{P}' \bar{\lambda}^2 \bar{\epsilon}\medsp
    &\quad   - \inv{4}
    \bar{P}'  (\bar{\chi} \gthree \gamma^a \bar{\lambda}) \bar{e}_a \wedge \ast
    \extd \bar{f} - \inv{32}\Bigl(\bar{P}''  - \half{1}
    \frac{\bigl[\bar{P}' \bigr]^2}{\bar{P}}\Bigr) \bar{\chi}^2 \bar{\lambda}^2 \bar{\epsilon} \biggr]
  \end{split}
\end{equation}

For the special case $\bar{Z} = 0$ discussed so far,
the identification \eqref{eq:genequiv} between the variables of MFS and of the
superfield formulation \eqref{eq:ps1} in \cite{Park:1993sd} becomes
trivial. Thus, replacing $\bar{\tve}$ in \eqref{eq:cmtransf} and
\eqref{eq:psicmtr} by $\bar{\ve}$,
the action \eqref{eq:mostgenaction2} with $\bar{Z} = 0$ together with
\eqref{eq:cmosh} is invariant under
\eqref{eq:elimtransf}-\eqref{eq:elimtransf3}, \eqref{eq:cmtransf} and \eqref{eq:psicmtr}. The
special case of minimal coupling ($P(\bar{\phi}) = 1$) of this matter extension of a gPSM based dilaton SUGRA
model has already been obtained in  ref.\
\cite{Izquierdo:1998hg} using
Noether techniques. But the above derivation using the equivalence of
this theory to a superspace formulation has definite advantages as it
straightforwardly generalizes to more complicated matter actions.

As a first step we should derive from the
result obtained so far the matter extension of MFS$_0$ (MFS$_0$ indicates MFS
for $\bar{Z} = 0$). This step is necessary as only the first
order formalism in terms
of a gPSM allows the straightforward treatment of the model at the
classical as well as at the quantum level. Also in the present context the MFS
formulation is much superior. The MFS action is different from
\eqref{eq:mostgenaction2}, which was obtained after
elimination of $X^a$ and (the torsion dependent part of) $\omega$. Now these auxiliary
variables must be re-introduced \emph{together} with the
matter coupling. Second, we would like to extend the matter coupling to the
most general MFS \eqref{eq:mostgenaction}. So far, we arrived at a matter
extension for the special
case $\bar{Z} = 0$ only. The general matter coupling ($Z \neq 0$) will be
obtained by the use of a certain dilaton dependent 
conformal transformation (cf.\ \cite{Ertl:2000si,Bergamin:2003am}). It is
argued in the end that the same result could also have been derived in a different way.

The discussion of a consistent matter extension of MFS$_0$ considerably
simplifies by observing that the action \eqref{eq:cmosh} or
\eqref{eq:cmdiffform} does not change when the independent variables $\bar{X}^a$ and
$\bar{\omega}$ are re-introduced. This is trivial for $\bar{\omega}$, as \eqref{eq:cmosh}
does not contain the dependent spin connection $\Tilde{\Bar{\omega}}$. The independence of $\bar{X}^a$ is obvious
as well \cite{Ertl:2000si,Bergamin:2003am}: The elimination condition of $\bar{X}^a$
depends on derivatives acting onto the dilaton field and the matter action
does not contain such terms. Thus the matter extension of MFS$_0$ must be of the
form\footnote{$X^I = (\phi, X^a, \chi^\alpha)$, $A_I = (\omega, e_a, \psi_\alpha)$}
\begin{equation}
  \label{eq:cmDPA}
  \Sbtext{tot}(\bar{X}, \bar{A}, \bar{f}, \bar{\lambda}) =
  \Stext{MFS$_0$}(\bar{X}, \bar{A}) + \Sbindex{(m)}(\bar{e}_a,
  \bar{\psi}_\alpha, \bar{\phi}, \bar{f}, \bar{\lambda}) \ .
\end{equation}
Not completely trivial is the derivation of the correct supersymmetry
transformations. However, it is important to realize that \eqref{eq:cmDPA}
already is invariant up
to equations of motion\footnote{We denote the equations of motion according to
the field which has been varied. Thus the $X^a$-e.o.m.\ refers to
\eqref{eq:gPSMeom4.2}, while the $\omega$-e.o.m.\ to
\eqref{eq:gPSMeom3.1}.} of $\bar{X}^a$ and $\bar{\omega}$: As
these e.o.m.-s are linear in $\bar{X}^a$ and $\bar{\omega}$, the
elimination of these fields ``commutes'' with the symmetry
transformation. Therefore there exists a simple and systematic way to modify
the MFS$_0$ symmetry laws \eqref{eq:gPSMtransf}-\eqref{eq:gPSMtransflast} such
that \eqref{eq:cmDPA} together with \eqref{eq:cmtransf} is again invariant. In
an abstract notation the behavior of \eqref{eq:cmDPA} under
\eqref{eq:gPSMtransf}-\eqref{eq:gPSMtransflast} and \eqref{eq:cmtransf} may be
written as
\begin{equation}
\begin{split}
  \label{eq:matternoninv}
  \bar{\delta} \Sbtext{tot}(\bar{X}, \bar{A}, \bar{f}, \bar{\lambda}) &= N_X^a(\bar{X},
  \bar{A}, \bar{f}, \bar{\lambda}; \bar{\ve}) \cdot (\mbox{$\bar{X}^a$-e.o.m.})
  \medsp &\quad  +
  (\mbox{$\bar{\omega}$-e.o.m.}) \wedge N_\omega (\bar{X},
  \bar{A}, \bar{f}, \bar{\lambda}; \bar{\ve})\ .
\end{split}
\end{equation}
Of course, the two field-dependent quantities $N_X^a$ and $N_\omega$
multiplying the e.o.m.-s must vanish in the absence of matter fields:
\begin{align}
   N_X^a(\bar{X},
  \bar{A}, \bar{f} = 0, \bar{\lambda} = 0; \bar{\ve}) &= 0\ , & N_\omega (\bar{X},
  \bar{A}, \bar{f} = 0, \bar{\lambda} = 0; \bar{\ve}) &= 0\ .
\end{align}
They are used to modify the symmetry
transformations of $\bar{X}^a$ and $\bar{\omega}$ by
\begin{align}
  \bar{\delta} \bar{X}^a &= \itindex{\bar{\delta}}{MFS$_0$} \bar{X}^a - N_X^a\ ,
  & \bar{\delta} \bar{\omega} &= \itindex{\bar{\delta}}{MFS$_0$} \bar{\omega} -
  N_\omega\ ,
\end{align}
where the transformations \eqref{eq:gPSMtransf}-\eqref{eq:gPSMtransflast} with
$\bar{Z} = 0$ have
been renamed $\itindex{\bar{\delta}}{MFS$_0$}$.

The explicit calculation of $N_X^a$ and $N_\omega$ is straightforward. As
$\Sbindex{(m)}$ depends on the MFS$_0$ fields $\bar{\phi}$, $\bar{e}_a$ and
$\bar{\psi}_\alpha$ only, the variations \eqref{eq:gPSMtransf}, \eqref{eq:gPSMtransf5} and
\eqref{eq:gPSMtransflast} within \eqref{eq:cmosh} lead to potential
non-invariance. But \eqref{eq:gPSMtransf} and \eqref{eq:gPSMtransf5} are equivalent to the supersymmetry
transformations of these fields within the superfield formulation (cf.\
\eqref{eq:howetransf}, \eqref{eq:pstransf}). There remains the supersymmetry transformation of the
gravitino. Again, most terms are equivalent to the superspace formulation
(remember that $\bar{Z} = 0$!), except for the covariant derivative $(D
\ve)_\alpha$. Here the dependent spin connection $\tilde{\omega}$ has been
replaced by the independent one. As the independent part of
the spin connection is eliminated by means of the $\bar{X}^a$-e.o.m.\
\eqref{eq:gPSMeom4.2} this leads to contributions to $N_X^a$. A further
source of non-invariance is the modification of the gravitino transformation
in \eqref{eq:psicmtr}. This yields another contribution to $N_X^a$ from the
covariant derivative acting on $\bar{\psi}$, but also to $N_\omega$ from the term
$\propto \bar{X}^a \bar{\psi} \gamma_a \bar{\psi}$. The latter contributions are proportional to
$P'(\bar{\phi})$ and vanish in the case of minimal coupling. Putting all terms
together one finds
\begin{align}
\begin{split}
  \label{eq:cmXtransf}
  \bar{\delta} \bar{X}^a &= \itindex{\bar{\delta}}{MFS$_0$} \bar{X}^a +
  \half{i} P  (\bar{\ve} \gamma^m \gamma^a
  \gthree \bar{\lambda}) \partial_m \bar{f} + \inv{4} P (\bar{\ve} \gamma^m \gamma^a \gthree
  \bar{\psi}_m) \bar{\lambda}^2\medsp
  &\quad - \frac{i}{16} P'  (\bar{\chi} \gamma^a \bar{\epsilon})
  \bar{\lambda}^2\ ,
\end{split}\medsp
  \label{eq:cmomegatransf}
  \bar{\delta} \bar{\omega} &= \itindex{\bar{\delta}}{MFS$_0$} \bar{\omega} -
  \inv{4} P' (\bar{\epsilon} \gthree \bar{\psi} - \bar{\epsilon}^\alpha \ast
  \bar{\psi}_\alpha) \bar{\lambda}^2\ .
\end{align}
With these new transformation laws the action \eqref{eq:cmDPA} is
finally fully invariant under supersymmetry, while local Lorentz invariance and
diffeomorphism invariance are manifest.
\subsection{Matter extension at $\mathbf{Z \neq 0}$\label{sec:5.2}}
To extend the matter couplings to the general MFS ($Z \neq 0$ in
\eqref{eq:mostgenaction}) we use the conformal transformation \eqref{eq:ct1}
and \eqref{eq:ct2} of sect.\ \ref{sec:susy}. The matter action is invariant
under those transformations of the fields, when the new matter fields are
defined as
\begin{align}
\label{eq:ctmatter}
  f &= \bar{f}\ , & \lambda &= e^{- \inv{4} Q(\phi)} \bar{\lambda}\ .
\end{align}
After the combined transformations
\eqref{eq:ct1}, \eqref{eq:ct2} and \eqref{eq:ctmatter} an action with
general MFS as geometrical part coupled to conformal matter is
obtained. $\Sbindex{(m)}$ in \eqref{eq:cmosh} by construction is invariant under
the conformal transformation and therefore that equation, after dropping all
bars, is the correct matter extension of MFS. Of course, the new action
\begin{equation}
 \label{eq:cmMFS}
   \Stext{tot}(X, A, f, \lambda) =
  \Stext{MFS}(X, A) + \Sindex{(m)}(e_a, \psi_\alpha, \phi, f, \lambda ) 
\end{equation}
is
invariant under the \emph{old} $\bar{\ve}$-transformations that act on the
barred variables
\begin{gather}
  \bar{\delta} \Stext{tot} \bigl(X(\bar{X}), A(\bar{A},\bar{X}), f(\bar{f}),
  \lambda(\bar{\lambda},\bar{X})\bigr) = 0\ ,
\end{gather}
but we have to be careful with the new transformations $\delta$ (depending on $\ve$), as the
transformation parameters themselves change under a conformal transformation as
well. The importance of this behavior for the
understanding of gPSM based SUGRA has been pointed out in
\cite{Bergamin:2003am}. Conformal transformations represent a special case of
target space diffeomorphisms in the (g)PSM formulation (cf.\ sect.\ 4.1 of
\cite{Bergamin:2003am}). Under such transformations the variables and symmetry
parameters change as
\begin{align}
\label{eq:transftransf}
  \bar{\delta} \bar{X}^I &= \delta \bar{X}^I (X)\ , \medsp
  \bar{\delta} \bar{A}_I &= \delta \bar{A}_I (A, X) + \mbox{e.o.m.-s}
  \label{eq:transftransf2}\ , \medsp
  \bar{\ve}_I &= \derfrac[X^J]{\bar{X}^I} \ve_J\ ,
\label{eq:transftransflast}
\end{align}
where the indices are the ones used in the gPSM formulation
\eqref{eq:gPSMaction}. Eq.\
\eqref{eq:transftransflast} together with \eqref{eq:ct1} and \eqref{eq:ct2}
for a pure supersymmetry transformation yield (cf.\ sect.\ 5.2 in ref.\
\cite{Bergamin:2003am}, esp.\ eq.\ (5.8)):
\begin{align}
\label{eq:susytransftransf}
  \bar{\ve} &= (\bar{\ve}_\phi, \bar{\ve}_a, \bar{\ve}_\alpha) = (0, 0, \bar{\ve}_\alpha)
  &\xrightarrow{\mbox{\tiny conformal transformation}} && \ve &= (\frac{Z}{4} (\chi \ve), 0, \ve_\alpha)
\end{align}
Thus for the general MFS the symmetries \eqref{eq:cmtransf} are modified by a local Lorentz
 transformation with field-dependent parameter $\ve_\phi = (1/4) Z \chi \ve$. The symmetry law of $f$
 remains unchanged under both, the conformal transformation \eqref{eq:ctmatter} and the additional
 local Lorentz transformation \eqref{eq:susytransftransf}, as this field is invariant under these
 symmetries. However, for $\lambda$ the local Lorentz transformation and the
 supersymmetry transformation of the conformal factor in \eqref{eq:ctmatter}
 add up to the new contributions displayed in eq.\ \eqref{eq:fintransflast} below.

Still the action \eqref{eq:cmMFS} is not invariant under
\eqref{eq:fintransf7}, \eqref{eq:fintransflast} and \eqref{eq:gPSMtransf}-\eqref{eq:gPSMtransflast}:
First, the modified laws of $\bar{\psi}$ \eqref{eq:psicmtr}, $\bar{X}^a$ \eqref{eq:cmXtransf}
and $\bar{\omega}$ \eqref{eq:cmomegatransf} should be rewritten in terms of the MFS
variables. But as none of these extensions generates derivatives onto the
conformal factors, this
boils down to rewrite these transformation rules in terms of variables without bars.
Second, the e.o.m.-s appearing on the r.h.s.\ of \eqref{eq:transftransf2}
may necessitate further modifications of the MFS symmetries. As the conformal
transformations \eqref{eq:ct1} and \eqref{eq:ct2} depend on the dilaton
field only, under supersymmetry
transformations discussed so far the action \eqref{eq:cmMFS} is invariant up to e.o.m.-s of
$\omega$. These new non-invariant terms originate from the variation of the
gravitino (cf.\ eq.\ (4.8) of ref.\ \cite{Bergamin:2003am} and comments below this equation). Indeed
a straightforward calculation shows that
\begin{align}
\label{eq:gravitinotransf}
  \bar{\delta} \psi_\alpha &= - \extd \bar{\epsilon}_\alpha
  + \inv{4} Z
  \extd \phi\, \bar{\epsilon}_\alpha + \ldots \ , & \delta
  \psi_\alpha &= - \extd \epsilon_\alpha + \ldots\ ,
\end{align}
where the dots indicate terms which do not contain derivatives.
The equation of motion of $\omega$ involved here drops out in the geometric
part, but obviously not in the matter extension. However, from the above
equation together with the formulation of $\Sindex{(m)}$ in
\eqref{eq:cmdiffform}, supersymmetry can be restored analogously to the
arguments leading to \eqref{eq:cmXtransf}. It can be read off from
\eqref{eq:gravitinotransf} that the
new (matter-field dependent) piece to the transformation of
$\omega$ is obtained by replacing $\psi_\alpha$ in
\eqref{eq:cmdiffform} by $1/4\, Z \epsilon_\alpha$:
\begin{equation}
  \label{eq:cmomegatransf2}
    \delta \omega = \mbox{\eqref{eq:cmomegatransf}} - 
    \frac{1}{4} Z P \Bigl( i (\ve \gamma^a \gamma^b \lambda) e_a^m \partial_m
    f (\ast e_b) + \inv{2} (\ve \gamma^a \gamma^b
    \psi_m) e_a^m (\ast e_b) \lambda^2 \Bigr)\ .
\end{equation}

It is useful to summarize what we have obtained in sect.\ \ref{se:matter}: The
gPSM based MFS models
of eq.\ \eqref{eq:mostgenaction} can be extended by the coupling of matter
fields. The complete action \eqref{eq:cmMFS} is given by the sum of
\eqref{eq:mostgenaction} and \eqref{eq:cmosh}.
The supersymmetry transformations \eqref{eq:gPSMtransf},
\eqref{eq:gPSMtransf3}, \eqref{eq:gPSMtransf5} for $\phi$, $\chi^\alpha$ and $e_a$  are not
changed by the matter coupling. Eq.\ \eqref{eq:gPSMtransflast} for the gravitino
receives new contributions
from the elimination of the auxiliary fields in superspace, while $\delta X^a$
and $\delta \omega$ are changed by re-introducing the auxiliary fields of the
gPSM formulation. Thus, the complete list of supersymmetry transformations is given 
by (\ref{eq:gPSMtransf})-(\ref{eq:gPSMtransflast}) plus new contributions from
the matter fields,
\begin{align}
\begin{split}
\label{eq:fintransf2}
\smallindex{\delta}{(m)} X^a &= \half{i} P (\ve \gamma^m \gamma^a
  \gthree \lambda) \partial_m f \medsp
 &\quad + \inv{4} P (\ve \gamma^m \gamma^a \gthree
  \psi_m) \lambda^2 - \frac{i}{16} P' (\chi \gamma^a \epsilon)
  \lambda^2 \ ,
\end{split}
\end{align}
\begin{align}
\begin{split}
\label{eq:fintransf4}
\smallindex{\delta}{(m)} \omega &= -
  \inv{4} P' (\epsilon \gthree \psi - \epsilon^\alpha \ast
  \psi_\alpha) \lambda^2 \medsp
 &\quad  - 
    \frac{1}{4} Z P \Bigl( i (\ve \gamma^a \gamma^b \lambda) e_a^m \partial_m
    f (\ast e_b) + \inv{2} (\ve \gamma^a \gamma^b
    \psi_m) e_a^m (\ast e_b)\lambda^2\Bigr)\ ,
\end{split}\medsp
\smallindex{\delta}{(m)} \psi_\alpha &= i \frac{P'}{8} \lambda^2 (\gamma^b
\ve)_\alpha e_b \ ,
\label{eq:fintransf6}
\end{align}
together with the transformations of the matter fields
\begin{align}
  \label{eq:fintransf7}
  \delta f &= i (\ve \lambda)\ ,\medsp
\begin{split}
  \delta \lambda_\alpha &=  \bigl( \partial_m f + i
  (\psi_m \lambda) \bigr) (\gamma^m \ve)_\alpha\medsp &\quad - \inv{8}
  Z \bigl( (\chi \gthree \ve) \lambda_\alpha + (\chi \ve)
  (\gthree \lambda)_\alpha \bigr) - \frac{P'}{4 P} (\lambda \gthree
  \chi) \ve_\alpha\ . \label{eq:fintransflast}
\end{split}
\end{align}

Of course, we could have used the relation to the general Park-Strominger model
SFDS of eq.\ \eqref{eq:ps3} to the MFS (cf.\
\eqref{eq:genequiv} and also ref.\ \cite{Bergamin:2003am}) instead of the conformal
transformation of MFS$_0$ to derive the matter coupling at $Z \neq
0$. Thus we may eliminate $X^a$ and $\omega$ in \eqref{eq:cmMFS} and by this
procedure arrive at the matter action \eqref{eq:cmosh} coupled to MFDS. Using
the techniques developed in \cite{Bergamin:2003am} this equivalence follows almost trivially. Considering the symmetry
transformations we note that we find again (cf.\ \eqref{eq:epsilonrel})
\begin{equation}
  \Delta \lambda_\alpha = - \inv{4} Z (\chi \ve) (\gthree \lambda)_\alpha
\end{equation}
in agreement with the result derived in \cite{Bergamin:2003am}.

One might wonder whether, on a different route, it is possible to derive a
different matter extension of gPSM based SUGRA, where modifications of
the transformation laws of $X^I$ and $A_I$ do not occur. The answer is
negative, as long as this extension shall preserve both, local Lorentz
invariance and supersymmetry. Indeed, the commutator of two local
supersymmetry transformations is a local Lorentz transformation $\delta_\phi$
plus a ``local translation'' $\delta_a$. Invariance under strict gPSM symmetry
transformations would imply that the matter action is invariant under
$\delta_a$, which, except for rigid supersymmetry, cannot be fulfilled.  

\section{Supersymmetric ground states with matter}
\label{se:mattergs}
The matter extension of MFS derived in the previous section allows the discussion of
supersymmetric ground states including matter fields. The fully supersymmetric
states are trivial: The geometric variables obey the same constraints
as derived already in section \ref{se:1}, the matter fields must obey $f =
\mbox{const.}$ and $\lambda = 0$.

More involved are the states with one supersymmetry: Eq.\ \eqref{eq:fintransf7}
leads to
\begin{equation}
  \label{eq:mgs1}
  \lambda^+ \ve_+ = - \lambda^- \ve_-
\end{equation}
which, in analogy to \eqref{sol:13}, implies $\lambda^2 \ve
\equiv 0$. Furthermore, as \eqref{eq:gPSMtransf}, \eqref{eq:gPSMtransf3} and
\eqref{eq:gPSMtransf5} did not receive new matter-field dependent
contributions, the relations \eqref{sol:13}, \eqref{eq:epscond1} and
\eqref{eq:epscond3} still hold. Of course, this still implies
\eqref{eq:epscond2}, but the geometric part of the Casimir function is no longer conserved (see
discussion below).

As a consequence of \eqref{eq:mgs1} and \eqref{sol:13}-\eqref{eq:epscond3}
the vanishing of $\delta \lambda_\alpha$ in \eqref{eq:fintransflast} reduces to
\begin{equation}
  \label{eq:mgs3}
  \delta \lambda_\alpha = (\gamma^m \ve)_\alpha \partial_m f = 0\ .
\end{equation}
It is straightforward to check that all matter-field dependent modifications
in \eqref{eq:fintransf2}-\eqref{eq:fintransf6} vanish due to \eqref{eq:mgs1}
and \eqref{eq:mgs3}. Thus the matter couplings do not change the
classification of the solutions as given in section \ref{se:2} as well as the
results of section \ref{se:3}.

In order to understand the condition \eqref{eq:mgs3} on the matter field
configurations it is advantagous to reformulate it as
\begin{align}
\label{eq:mgs31}
  f^{++} \ve_+ &= 0\ , & f^{--} \ve_- &= 0\ ,
\end{align}
where $f^{\pm \pm} = \ast (e^{\pm \pm} \wedge \extd f)$ are
(anti-)selfdual field configurations of the scalar field. Thus the chiral solution of
\eqref{sol:13} and \eqref{eq:mgs1} with $\chi^- = \lambda^- = \ve_+ = 0$
admits selfdual scalar fields while the anti-chiral one allows anti-selfdual
$f$. The third solution with $\ve_+ \neq 0$ and $\ve_- \neq 0$ is compatible
with static $f$, only.

Even in the presence of matter a conserved quantity can be constructed.
Its physical relevance is displayed in the close
relationship to energy definitions well-known from General Relativity,
such as ADM-, Bondi- and quasi-local mass (for details we refer to sect.\ 5 of
\cite{Grumiller:2002nm} and references therein).

The conservation law $\extd C = 0 $ in the presence of matter fields is
modified by analogy to the pure bosonic case (cf.\ \cite{Grumiller:1999rz}) 
according to 
\begin{gather}
  \label{eq:mgs4}
  e^{-Q} \extd C + X^{--} W^{++} + X^{++} W^{--} + \inv{8}(u' + \half{1} u Z)
  (\chi^- W_- - \chi^+ W_+)  = 0 \medsp\label{eq:mgs5}
\begin{alignat}{2}
  W^{\pm \pm} &= \varfrac{e^{\mp \mp}} \Sindex{(m)} {} &\qquad \quad W_{\pm} &= \varfrac{\psi^{\mp}} \Sindex{(m)}
\end{alignat}
\end{gather}
In the presence of matter $(- W^{\pm \pm})$ appear on the r.h.s.\ of the
e.o.m-s \eqref{eq:gPSMeom3.2}, $(- W_{\pm})$ on the r.h.s.\ of
\eqref{eq:gPSMeom3.3}. Eq.\ \eqref{eq:mgs4} results from a suitable linear
combination of the e.o.m-s \eqref{eq:gPSMeom3.1}-\eqref{eq:gPSMeom3.3}, $C$ now is only \emph{part} of a total conserved
quantity, which also contains a matter contribution $e^{-Q} \extd
\smallindex{C}{(m)}$ from the $W$ terms.

A straightforward calculation from eq.\ \eqref{eq:cmosh} yields
\begin{align}
  \begin{split}
    \label{eq:mgs5a}
    W^{\pm \pm} &= \pm e^{\pm \pm} \biggl[ P \Bigl( f^{++} f^{--} + i \ast
    (e^{++} \psi \lambda) f^{--} + i \ast
    (e^{--} \psi \lambda) f^{++} +\medsp 
    &\quad  \half{1} \ast (\psi e^{--})^\alpha \ast
    (\psi e^{++})_\alpha (\lambda \lambda)\Bigr)
    + \frac{u}{8} P' \lambda^2 +\inv{32} \bigl(P'' - \half{1} \frac{{P'}^2}{P} \bigr) \chi^2
    \lambda^2 \biggr]\medsp
    &\quad + \extd f \Bigl[ P \bigl( f^{\pm \pm} + i \ast (e^{\pm
    \pm} \psi \lambda) \bigr) - \frac{i}{2 \sqrt{2}} P' \chi^{\pm}
    \lambda^{\pm}\Bigr]\medsp
    &\quad + P \Bigl[ i (\psi \lambda) f^{\pm \pm} - \half{1} \psi \ast(\psi
    e^{\pm \pm}) (\lambda \lambda) \mp \inv{\sqrt{2}} \lambda^{\pm} \extd
    \lambda^{\pm} \Bigr] \ ,
  \end{split}
\end{align}
\begin{align}
  \begin{split}
    \label{eq:mgs6}
    W_{\mp} &= P \biggl[ -i \lambda^{\pm} (e^{++} f^{--} + e^{--} f^{++} \pm
    \extd f) - \half{1} \psi_{\mp} (\lambda \lambda) \medsp
    &\quad \pm \half{1} \bigl(e^{++} \ast(\psi_{\mp} e^{--}) + e^{--}
    \ast(\psi_{\mp} e^{++}) \bigr) (\lambda \lambda) \biggr]\ .
  \end{split}
\end{align}

Now also the question may be posed about the meaning of the restriction
\eqref{eq:epscond2} within that generalized conservation law.
The body of $\extd C$ in \eqref{eq:mgs4} vanishes trivially due to that equation. The restriction to (anti-)selfdual or static $f$ as
derived from \eqref{eq:mgs31} ensures that the body of \eqref{eq:mgs4}
vanishes without imposing further constraints on the fields. Thus the result
of sect.\ \ref{se:3} for the matterless case continues to hold if non-minimally
coupled conformal matter is included.
\section{Conclusions}
In our present work we present the complete classification of all BPS BHs in
2D dilaton SUGRA coupled to conformal matter. The use of a first order
formulation as suggested from the graded Poisson-Sigma model approach for the geometric part of the action plays a crucial r\^{o}le in
the calculations. As no matter extension thereof had been considered in the literature,
its derivation is an important result on its own. For future reference we
compile the MFS action non-minimally coupled to conformal matter (with
coupling function $P(\phi)$) at this
place:
\begin{multline}
\label{eq:MFS+cm}
      \mathcal{S} = \int_{\mathcal{M}} \Biggl[ \phi \diff \omega + X^a D e_a + \chi^\alpha D
  \psi_\alpha + \epsilon \biggl( V + Y Z - \half{1} \chi^2 \Bigl( \frac{VZ + V'}{2u} +
  \frac{2 V^2}{u^3} \Bigr) \biggr) \medsp
   + \frac{Z}{4} X^a
    (\chi \gamma_a \gamma^b e_b \gthree \psi) + \frac{i V}{u}
  (\chi \gamma^a e_a \psi)
   + i X^a (\psi \gamma_a \psi) - \half{1} \bigl( u +
  \frac{Z}{8} \chi^2 \bigr) (\psi \gthree \psi) \medsp
  + {P}  \Bigl( \half{1} \extd {f} \wedge
    \ast \extd {f} + \half{i}
    {\lambda} \gamma_a {e}^a \wedge \ast \extd {\lambda} + i \ast({e}_a \wedge \ast \extd {f}) {e}_b \wedge \ast {\psi} \gamma^a
    \gamma^b {\lambda} \medsp
    \quad + \inv{4} \ast ({e}_b \wedge \ast {\psi}) \gamma^a \gamma^b
    {e}_a \wedge \ast {\psi} {\lambda}^2 \Bigr) +
    \frac{{u}}{8} {P}' {\lambda}^2 {\epsilon}\medsp
    \quad   - \inv{4}
    {P}'  ({\chi} \gthree \gamma^a {\lambda}) {e}_a \wedge \ast
    \extd {f} - \inv{32}\Bigl({P}''  - \half{1}
    \frac{\bigl[{P}' \bigr]^2}{{P}}\Bigr) {\chi}^2 {\lambda}^2 {\epsilon} \Biggr]
\end{multline}
This action is invariant under the supersymmetry transformations
\eqref{eq:gPSMtransf}-\eqref{eq:gPSMtransflast} supplemented by
\eqref{eq:fintransf2}-\eqref{eq:fintransflast} and \eqref{eq:fintransf7},\eqref{eq:fintransflast}.

Starting from this action it has been shown that all BPS like states
have vanishing body of the Casimir function and thus are
ground states. Solutions with vanishing fermions allow a non-trivial bosonic
geometry, but all Killing horizons were found to be extremal. On the other
hand, the geometry of solutions with
non-vanishing fermions must be Minkowski space and consequently there exist no
supersymmetric BHs with dilatino or gravitino hair. The impossibility of
supersymmetric dS ground states has been reproduced for our class of models
and the absolute conservation law---the modification of the Casimir function
in presence of matter fields---has been calculated explicitely. 

\paragraph{Note added in proofs:} 
While proof reading an e-print appeared \cite{Davis:2004xb} which allows a nice
application of some of the current paper's methods. That study is based
upon 2D type 0A string theory and among other issues an upper bound on the
number $q\leq 16\pi e < 12$ of electric and magnetic D0 branes is derived
(eq. (4.7) of \cite{Davis:2004xb}). The same bound immediately follows from reality of
the prepotential
\[
u(\phi)\propto \sqrt{1-(q^2/(16\pi))(\ln{\phi}/\phi)}
\]
or, equivalently, from semi-negativity of $w(\phi)$ in (B.12). Note that
our dilaton $\phi$ is related to the dilaton $\Phi$ in \cite{Davis:2004xb} by
$\phi=\exp{(-2\Phi)}$. In addition, as a simple consequence of the
conservation of the Casimir function (B.9) we agree on the result for the
ADM mass (eq. (3.9) of \cite{Davis:2004xb}).

\section*{Acknowledgement}

This work has been supported by projects P-14650-TPH and P-16030-N08 of the
Austrian Science Foundation (FWF). We thank P.\ van Nieuwenhuizen and Th.\
Mohaupt for helpful
discussions on dS vacua and BPS like
black holes, resp. We are grateful to D. Vassilevich and V. Frolov for asking
very relevant questions. This work has been completed in the hospitable
atmosphere of the International Erwin Schr\"{o}dinger Institute. We would like
to thank one of the referees for suggesting important improvements.

%%
%%APPENDIX
%%
\appendix
\section{Notations and conventions}
\label{sec:notation}
These conventions are identical to
\cite{Ertl:2000si,Ertl:2001sj}, where additional explanations can be found.

Indices chosen from the Latin alphabet are commuting (lower case) or
generic (upper case), Greek indices are anti-commuting. Holonomic coordinates
are labeled by $M$, $N$, $O$ etc., anholonomic ones by $A$, $B$, $C$ etc., whereas
$I$, $J$, $K$ etc.\ are general indices of the gPSM. The index $\phi$ is used to indicate the dilaton component of
  the gPSM fields:
  \begin{align}
    X^\phi &= \phi & A_\phi &= \omega
  \end{align}

The summation convention is always $NW \rightarrow SE$, e.g.\ for a
fermion $\chi$: $\chi^2 = \chi^\alpha \chi_\alpha$. Our conventions are
arranged in such a way that almost every bosonic expression is transformed
trivially to the graded case when using this summation convention and
replacing commuting indices by general ones. This is possible together with
exterior derivatives acting \emph{from the right}, only. Thus the graded
Leibniz rule is given by
\begin{equation}
  \label{eq:leibniz}
  \mbox{d}\left( AB\right) =A\mbox{d}B+\left( -1\right) ^{B}(\mbox{d}A) B\ .
\end{equation}

In terms of anholonomic indices the metric and the symplectic $2 \times 2$
tensor are defined as
\begin{align}
  \eta_{ab} &= \left( \begin{array}{cc} 1 & 0 \\ 0 & -1
  \end{array} \right)\ , &
  \epsilon_{ab} &= - \epsilon^{ab} = \left( \begin{array}{cc} 0 & 1 \\ -1 & 0
  \end{array} \right)\ , & \epsilon_{\alpha \beta} &= \epsilon^{\alpha \beta} = \left( \begin{array}{cc} 0 & 1 \\ -1 & 0
  \end{array} \right)\ .
\end{align}
The metric in terms of holonomic indices is obtained by $g_{mn} = e_n^b e_m^a
\eta_{ab}$ and for the determinant the standard expression $e = \det e_m^a =
\sqrt{- \det g_{mn}}$ is used. The volume form reads $\epsilon = \half{1}
\epsilon^{ab} e_b \wedge e_a$; by definition $\ast \epsilon = 1$.

The $\gamma$-matrices are used in a chiral representation:
\begin{align}
\label{eq:gammadef}
  {{\gamma^0}_\alpha}^\beta &= \left( \begin{array}{cc} 0 & 1 \\ 1 & 0
  \end{array} \right) & {{\gamma^1}_\alpha}^\beta &= \left( \begin{array}{cc} 0 & 1 \\ -1 & 0
  \end{array} \right) & {{\gthree}_\alpha}^\beta &= {(\gamma^1
    \gamma^0)_\alpha}^\beta = \left( \begin{array}{cc} 1 & 0 \\ 0 & -1
  \end{array} \right)
\end{align}

Covariant derivatives of anholonomic indices with respect to the geometric
variables $e_a = \extd x^m e_{am}$ and $\psi_\alpha = \extd x^m \psi_{\alpha m}$
include the two-dimensional spin-connection one form $\omega^{ab} = \omega
\epsilon^{ab}$. When acting on lower indices the explicit expressions read
($\half{1} \gthree$ is the generator of Lorentz transformations in spinor space):
\begin{align}
\label{eq:A8}
  (D e)_a &= \extd e_a + \omega {\epsilon_a}^b e_b & (D \psi)_\alpha &= \extd
  \psi_\alpha - \half{1} {{\omega \gthree}_\alpha}^\beta \psi_\beta
\end{align}

Light-cone components are very convenient. As we work with spinors in a
chiral representation we can use
\begin{align}
\label{eq:Achi}
  \chi^\alpha &= ( \chi^+, \chi^-)\ , & \chi_\alpha &= \begin{pmatrix} \chi_+ \\
  \chi_- \end{pmatrix}\ .
\end{align}
For Majorana spinors upper and lower chiral components are related by $\chi^+
= \chi_-$, $ \chi^- = - \chi_+$, $\chi^2 = \chi^\alpha \chi_\alpha = 2 \chi_- \chi_+$. Vectors in light-cone coordinates are given by
\begin{align}
\label{eq:A10}
  v^{++} &= \frac{i}{\sqrt{2}} (v^0 + v^1)\ , & v^{--} &= \frac{-i}{\sqrt{2}}
  (v^0 - v^1)\ .
\end{align}
The additional factor $i$ in \eqref{eq:A10} permits a direct identification of the light-cone components with
the components of the spin-tensor $v^{\alpha \beta} = \frac{i}{\sqrt{2}} v^c \gamma_c^{\alpha
  \beta}$. This implies that $\eta_{++|--} = 1$
and $\epsilon_{--|++} = - \epsilon_{++|--} = 1$.
\section{E.o.m.-s and conserved quantity}\label{sec:eom}
The equations of motion for a generic gPSM are
\begin{align}
\label{eq:gPSMeom1}
  \extd X^I + P^{IJ} A_J &= 0\ ,\medsp
\label{eq:gPSMeom2}
  \extd A_I + \half{1} (\partial_I P^{JK}) A_K A_J &= 0\ .
\end{align}
Consequently, the ones for the MFS action \eqref{eq:mostgenaction} become:
\begin{gather}
  \extd \phi - X^b {\epsilon_b}^a e_a + \half{1} (\chi \gthree \psi ) = 0 \label{eq:gPSMeom3.1}
  \medsp
  D X^a + \epsilon^{ab} e_b \biggl( V + Y Z - \half{1} \chi^2 \Bigl( \frac{VZ + V'}{2u} +
  \frac{2 V^2}{u^3} \Bigr) \biggr)\hspace{5cm}  \nonumber \medsp \hspace{5cm} - \frac{Z}{4} X^b
    (\chi \gamma_b \gamma^a \gthree \psi) - \frac{i V}{u}
  (\chi \gamma^a \psi) = 0 \label{eq:gPSMeom3.2}
\end{gather}
\begin{gather}
  D \chi^\alpha + \frac{Z}{4} X^a
    {(\chi \gamma_a \gamma^b \gthree)}^\alpha e_b + \frac{i V}{u}
  (\chi \gamma^a)^\alpha e_a \hspace{5cm} \nonumber \medsp \hspace{5cm} + 2 i X^a (\psi \gamma_a)^\alpha - \bigl( u +
  \frac{Z}{8} \chi^2 \bigr) (\psi\gthree)^\alpha  = 0 \label{eq:gPSMeom3.3}\medsp
  \extd \omega + \epsilon \biggl( V' + Y Z' - \half{1} \chi^2 \Bigl( \bigl(\frac{VZ + V'}{2u}\bigr)' +
  \bigl(\frac{2 V^2}{u^3}\bigr)' \Bigr) \biggr)\hspace{5cm}\nonumber \medsp + \frac{Z'}{4} X^b
    (\chi \gamma_b \gamma^a e_a \gthree \psi) 
  +  i \bigl(\frac{V}{u}\bigl)'
  (\chi \gamma^a e_a \psi) - \half{1} \bigl( u' +
  \frac{Z'}{8} \chi^2 \bigr) (\psi \gthree \psi) = 0 \label{eq:gPSMeom4.1}\medsp
  D e_a + \eta_{ab} \epsilon X^b Z + \frac{Z}{4}
    (\chi \gamma_a \gamma^b e_b \gthree \psi) + i (\psi \gamma_a \psi) = 0 \label{eq:gPSMeom4.2} 
  \medsp
  D \psi_\alpha - \epsilon \chi_\alpha \Bigl( \frac{VZ + V'}{2u} +
  \frac{2 V^2}{u^3} \Bigr) + \frac{Z}{4} X^a
    (\gamma_a \gamma^b e_b \gthree\psi)_\alpha +\hspace{5cm} \nonumber \medsp \hspace{5cm} \frac{i V}{u}
  (\gamma^a e_a \psi)_\alpha - \frac{Z}{8} \chi_\alpha (\psi \gthree \psi) = 0
  \label{eq:gPSMeom4.3} 
\end{gather}
We re-emphasize that $V,Z$ and the prepotential $u$ are related by \eqref{eq:sol1}.

The full analytic solution of MFS has been given in sect.\ 6 of
\cite{Bergamin:2003am}. Each solution is characterized by a certain value of
the Casimir function, a quantity conserved in space and time. It consists of a
bosonic part (body) and a fermionic one (soul):
\begin{align}
\label{sol:13.5}
C &= C_B + C_S \medsp
\label{sol:14}
C_B &=e^{Q(\phi)}Y+w(\phi)= e^{Q(\phi)}\left(Y-\frac 18
  u^2(\phi)\right) \medsp
\label{sol:12}
C_S &= \inv{16} e^Q \chi^2 (u' + \half{1} u Z)
\end{align}
In this equation  the (logarithm of the) integrating factor and the conformally invariant combination of the bosonic potentials
\begin{align}
\label{eq:sol2}
Q(\phi)&:=\int^\phi Z(\phi')d\phi'\ , &  w(\phi)&:=\int^\phi
e^{Q(\phi')}V(\phi')d\phi'=-\frac 18 e^{Q(\phi)}u^2(\phi) \leq 0
\end{align}
have been introduced. In \cite{Bergamin:2003am} the solutions for $C \neq 0$
(eqs.\ (6.9)-(6.13)) and $C = 0$ (eqs.\ (6.17)-(6.21)) have been given which
are not reproduced here. 
\section{Dilaton SUGRA in superspace}
\label{sec:integrals}
The action for a general dilaton SUGRA in superspace \cite{Park:1993sd} may be written as\footnote{In
  ref.\ \cite{Park:1993sd} the first term in the brackets was chosen as $E J(\Phi) S$. If a
  global field redefinition $J(\Phi) \rightarrow \Phi$ is not possible, these
  models are not equivalent globally to MFS
  \cite{Bergamin:2003am}.}
\begin{equation}
  \label{eq:ps1}
  \Stext{SFDS} = \intd{\diff{^2x} \diff{^2 \theta}} E \bigl( \Phi S -
  \inv{4} Z(\Phi) D^\alpha \Phi D_\alpha \Phi + \half{1} u(\Phi) \bigl)\ ,
\end{equation}
with\footnote{Except for the zweibein, components of superfields are denoted by underlined
  variables to distinguish them from the fields in the gPSM approach.} \cite{Howe:1979ia}
\begin{align}
\label{eq:superdet}
  E &= e \bigl( 1 + i \theta \gamma^a \underline{\psi}_a + \half{1} \theta^2
  ( \underline{A} + \epsilon^{ab} \underline{\psi}_b \gthree
  \underline{\psi}_a )\ , \medsp
\begin{split}
  S &= \underline{A} + 2 \theta \gthree \tilde{\underline{\sigma}} - i
  \underline{A} \theta \gamma^a \underline{\psi}_a\medsp &\quad + \half{1}
  \theta^2 \bigl( \epsilon^{mn} \partial_n \tilde{\underline{\omega}}_m - \underline{A} (\underline{A} + \epsilon^{ab} \underline{\psi}_b \gthree
  \underline{\psi}_a ) - 2i \underline{\psi}^a \gamma_a \gthree
  \tilde{\underline{\sigma}} \bigl)\ ,
\end{split} \medsp
\label{eq:susycovder}
D_\alpha &= \partial_\alpha + i (\gamma^a \theta)_\alpha \partial_a\ .
\end{align}
Quantities with a tilde are defined  in analogy to footnote \ref{fn:tildedef}.
$\Phi$ is the
dilaton superfield with component expansion
\begin{equation}
  \label{eq:dilatonsf}
  \Phi = \underline{\phi} + \half{1} \theta \gthree \underline{\chi} + \half{1} \theta^2 \underline{F}\ .
\end{equation}
The supersymmetry transformations of the component fields of the
superdeterminant are given by
\begin{gather}
\label{eq:howetransf}
  \begin{alignat}{2}
    \delta {e_m}^a &= - 2i (\tve \gamma^a \underline{\psi}_m)\ , &\qquad \delta {e^m}_a &= 2
    i (\tve \gamma^m \underline{\psi}_a)\ ,
  \end{alignat} \medsp
\label{eq:howetransf2}
    \delta {\underline{\psi}_m}^\alpha = - \bigl( (\tilde{\underline{D}} \tve)^\alpha + \half{i} \underline{A}
    (\tve \gamma_m)^\alpha\bigr)\ , \medsp
\label{eq:howetransflast}
    \delta \underline{A} = -2 \bigl( (\tve \gthree \tilde{\underline{\sigma}}) - \half{i} \underline{A} {e^m}_a
    (\tve \gamma^a \underline{\psi}_m) \bigr)\ ,
\end{gather}
while the ones of the dilaton superfield read:
\begin{align}
\label{eq:pstransf}
  \delta \underline{\phi} &= - \half{1} \tve \gthree \underline{\chi} \medsp
\label{eq:pstransf2}
  \delta \underline{\chi}_\alpha &= - 2 (\gthree \tve)_\alpha \underline{F} + i (\gthree \gamma^b
  \tve)_\alpha (\underline{\psi}_b \gthree \underline{\chi}) - 2 i (\gthree \gamma^m \tve)_\alpha
  \partial_m \underline{\phi} \medsp
\label{eq:pstransflast}
  \delta \underline{F} &= i (\tve \gamma^a \underline{\psi}_a) \underline{F} - \half{i} \bigl(\tve \gamma^m \gthree
  (\tilde{\underline{D}}_m \underline{\chi}) \bigr) + (\tve \underline{\lambda}^m) \bigl((\underline{\psi}_m \gthree \underline{\chi}) - 2
  \partial_m \underline{\phi} \bigr)
\end{align}
Integrating out superspace and eliminating the auxiliary fields
$\underline{A}$ and $\underline{F}$ using their equations of motion
\begin{align}
\label{eq:psauxfield}
  \underline{F} &= - \half{u} \ , & \underline{A} &=
   -\half{1} ( u' + u Z) + \inv{8}
   Z' \underline{\chi}^2 \ . 
\end{align}
one arrives at the action
\begin{equation}
  \label{eq:ps3}
  \begin{split}
    \Stext{SFDS} &= \intd{\diff{^2x}} e \biggl( \half{1} \underline{\tilde{R}} \underline{\phi} +
    (\underline{\chi} \tilde{\underline{\sigma}}) - \half{1} Z \bigl(\partial^m \underline{\phi} \partial_m \underline{\phi} - \frac{i}{4} \underline{\chi} \gamma^m
  \partial_m \underline{\chi} \medsp
&\phantom{= \intd{\diff{^2x}} e \biggl(}  - (\underline{\psi}_n \gamma^m
    \gamma^n \gthree \underline{\chi}) \partial_m \underline{\phi} \bigr) -
    \inv{8}\bigl((u^2)' + u^2 Z \bigr) + \half{u} \epsilon^{mn}
    (\underline{\psi}_n \gthree \underline{\psi}_m) \medsp
    &\phantom{= \intd{\diff{^2x}} e \biggl(} + \half{i} u' (\underline{\zeta} \gthree \underline{\chi})
     + \inv{8} \bigl( u'' + \inv{4} u
    Z'  + \half{1} Z (\underline{\psi}_n \gamma^m \gamma^n
    \underline{\psi}_m) \bigr)
    (\underline{\chi}\underline{\chi}) \biggl)\ ,
  \end{split}
\end{equation}
while the symmetry transformations of the remaining fields take the form
\begin{gather}
\label{eq:psostransf}
\begin{alignat}{2}
    \delta {e_m}^a &= - 2i (\tve \gamma^a \underline{\psi}_m)\ , &\qquad \delta {e^m}_a &= 2
    i (\tve \gamma^m \underline{\psi}_a)\ ,
\end{alignat}\medsp
\label{eq:psostransf2}
    \delta {\underline{\psi}_m}^\alpha = - (\tilde{D} \tve)^\alpha +
    \frac{i}{4} \bigl( u' + u Z - \inv{8} Z'  (\underline{\chi}\underline{\chi})\bigr)  
    (\tve \gamma_m)^\alpha\ , \medsp 
\label{eq:psostransf3}
  \delta \underline{\phi} = - \half{1} \tve \gthree \underline{\chi}\ , \medsp
\label{eq:psostransflast}
  \delta \underline{\chi}_\alpha =  u (\gthree \tve)_\alpha  + i (\gthree \gamma^b
  \tve)_\alpha (\underline{\psi}_b \gthree \underline{\chi}) - 2 i (\gthree \gamma^m \tve)_\alpha
  \partial_m \underline{\phi}\ .  
\end{gather}
In ref.\ \cite{Bergamin:2003am} it has been shown that this action is
equivalent to the action \eqref{eq:mostgenaction2} of MFDS if the identifications
\begin{align}
\label{eq:genequiv}
  \underline{\psi}_m^\alpha &= \psi_m^\alpha + \frac{i}{8} Z(\phi) e^a_m
  \epsilon_{ab} (\chi \gamma^b)^\alpha\ , &\underline{\phi} &= \phi\ , & \underline{\chi} &= \chi\ , 
\end{align}
are made. The supersymmetry transformations are equivalent up to a local
Lorentz transformation with field dependent parameter:
\begin{align}
\label{eq:epsilonrel}
  \ve &= \tve & \Delta &= \itindex{\delta}{MFDS} - \itindex{\delta}{SFDS} =
  \half{Z} \chi \ve\, \delta_{\phi}
\end{align}
\providecommand{\href}[2]{#2}\begingroup\raggedright\endgroup

\end{document}